\documentclass[aps,prx,twocolumn,footinbib,showkeys,nobalancelastpage]{revtex4-2}

\usepackage{times}
\usepackage{latexsym}
\usepackage{graphicx}
\usepackage{amsmath,amssymb}
\usepackage{tensor}
\usepackage{verbatim,bbm}
\usepackage{textcomp}

\usepackage{indentfirst}
\usepackage{physics}
\usepackage{braket}
\usepackage{float}
\usepackage{upgreek}
\usepackage{verbatim}
\usepackage{epstopdf}
\usepackage{CJK}
\usepackage{esint}
\usepackage{color}
\usepackage[T1]{fontenc}
\usepackage{subfigure}
\usepackage{amsfonts}
\usepackage{footmisc}
\usepackage{scrextend}
\usepackage{multirow}
\usepackage[hyperfootnotes=false]{hyperref}

\usepackage[english]{babel}
\usepackage{url}
\usepackage{bm}

\usepackage{url}
\usepackage{bm}
\usepackage{hyperref}
\definecolor{darkblue}{rgb}{0,0,0.5}
\hypersetup{
colorlinks=true,
linkcolor=black,
filecolor=blue,
citecolor=darkblue,  
urlcolor=black,
}

\begin{document}

\title{Optimal environment localization}
\author{Jason L. Pereira$^{1}$}
\author{Quntao Zhuang$^{2,3}$}
\author{Stefano Pirandola$^{1}$}
\affiliation{
$^1$ Department of Computer Science, University of York, York YO10 5GH, UK
\\
$^2$ Department of Electrical and Computer Engineering, University of Arizona, Tucson, AZ 85721, USA
\\
$^3$ James C. Wyant College of Optical Sciences, University of Arizona, Tucson, AZ 85721, USA
}
\date{\today}

\begin{abstract}
Quantum channels model many physical processes. For this reason, hypothesis testing between quantum channels is a fundamental task in quantum information theory. Here we consider the paradigmatic case of channel position finding, where the aim is to determine the position of a target quantum channel within a sequence of background channels. We explore this model in the setting of bosonic systems, considering Gaussian channels with the same transmissivity (or gain) but different levels of environmental noise. Thus the goal of the problem becomes detecting the position of a target environment among a number of identical background environments, all acting on an input multi-mode system. We derive bounds for the ultimate error probability affecting this multi-ary discrimination problem and find an analytic condition for quantum advantage over protocols involving classical input states. We also design an explicit protocol that gives numerical bounds on the ultimate error probability and often achieves quantum advantage. Finally, we consider direct applications of the model for tasks of thermal imaging (finding a warmer pixel in a colder background) and quantum communication (for localizing a different level of noise in a sequence of lines or a frequency spectrum).
\end{abstract}

\maketitle

\section{Introduction}
Quantum channel discrimination (QCD)~\cite{KitaevDiamond,Acin_2001,acin2001statistical,sacchi2005entanglement,Sacchi_2005_2,wang2006unambiguous} is an important task in quantum computing~\cite{Shor_1997,linear_optics_non_universal} and quantum communication~\cite{Bennett20147,Bennett2002,shipractical}. Quantum channels model the input and output relation of quantum states in physical processes~\cite{Nielsen2002,hayashi2006quantum,Holevo_2011}. Various applications in quantum sensing~\cite{pirandola2018advances} can be reduced to QCD problems. An important case of QCD is finding a target channel within a sequence of background channels. 
~In this case, we have a sequence of channels and know that all but one of them (the background channels) are identical, whilst one of them (the target channel) is different. The goal is to figure out which channel is the target channel, by probing the sequence of channels with quantum states a set number of times. This is a task of channel-position finding (CPF) \cite{zhuang_entanglement-enhanced_2020}.

It is important to note that there are relevant scenarios in quantum sensing where the transmissivity stays the same for all channels while the noise background differs. This is a scenario with a passive signature, meaning that different levels of noise can be detected at the output of the channels even in the absence of input signals. In this setting, the model of CPF becomes a problem of environment localization, where the aim is to optimally identify the position of a different (target) environment with respect to standard background environments affecting an ensemble of modes. Motivated by this observation, we study CPF among bosonic Gaussian channels~\cite{weedbrook2012gaussian} with the same transmissivity (or gain) but different environments, establish the ultimate performance of this problem, and identify the regime of parameters where we can have quantum advantage over the classical benchmark based on coherent states.

More precisely, we use channel simulation and stretching techniques~\cite{pirandola_fundamental_2017,pirandola2017ultimate,pirandola2019fundamental} to find the minimum fidelity between the outputs of two Gaussian channels that have the same transmissivity, $\tau$, but give rise to different induced noises, $\nu$. This minimization is carried out over all quantum inputs. We then use this minimum fidelity to find upper and lower bounds on the minimum discrimination error in finding the position of a target channel within a sequence of channels, for a fixed number of probes sent through each channel of the sequence. These bounds are on the minimum discrimination error for all possible adaptive, quantum protocols. We also find the minimum fidelity between two channel outputs (for channels with the same value of $\tau$ but different values of $\nu$) where the minimization is carried out over classical input states (mixtures of coherent states). We use this fidelity to find a lower bound on the minimum discrimination error for all possible classical protocols.

Our quantum and classical bounds hold for all phase-insensitive, Gaussian channels (thermal loss channels, thermal amplifier channels and additive noise channels)~\cite{holevo2007one,weedbrook2012gaussian}. By comparing these bounds, we are able to prove quantum advantage for the general problem of environment localization. In particular, we find a condition on the sequence of channels that guarantees quantum advantage if the number of probes sent through the sequence of channels is large enough. Furthermore, we also design an explicit protocol, based on entangled states, photon counting, and the maximum-likelihood estimation, which is able to beat any classical strategy. We apply our bounds (and the explicit protocol) to a number of discrimination tasks. We consider thermal imaging to find a warmer pixel in a colder background, eavesdropper localization to find the channel that an eavesdropper is interfering with and the problem of finding the least noisy frequency in a multi-mode cable.

\section{Results}
Our main results are upper and lower bounds on the error probability of environment localization. To establish the bounds, in Section~\ref{sec:channel simulation}, we present a method of channel simulation which allows the reduction of arbitrary adaptive protocols to quantum operations on a sequence of Choi states. From there, fidelity-based bounds can be derived for the error probability, which are calculated explicitly for Gaussian channels in Section~\ref{sec:fidelity}. In Section~\ref{sec:classical limits}, we use similar techniques to bound the error of classical protocols. In Section~\ref{sec:quantum advantage}, we then establish a region in which we can analytically prove that the task shows a quantum advantage. We present a concrete receiver design and discrimination protocol in Section~\ref{sec: bounds protocols} and thereby provide numerical bounds on the error, in both the classical and quantum cases. These bounds are often tighter than our fidelity-based, analytic bounds, and so we are often able to demonstrate quantum advantage at a lower number of probes than is required for our analytic bounds. Finally, we apply our bounds to several examples, in Section~\ref{sec:applications}, and demonstrate the quantum advantage.

\subsection{Channel simulation}
\label{sec:channel simulation}
Consider a sequence of $m$ one-mode, phase-insensitive, Gaussian channels, where $m-1$ of the channels are identical ``background'' channels and one of the channels is a target channel. The target channel has the same transmissivity, $\tau$, as the background channels, but a different induced noise, $\nu$ (note that we consider a generalized transmissivity which may take values between zero and infinity). Suppose we want to identify the target channel and can do so by probing the sequence of channels using some adaptive protocol that involves sending $M$ transmissions through the sequence of channels (each transmission consists of sending a one-mode state through every channel in the sequence). We do not impose any energy bound on the transmissions. We would like to bound the minimum probability of error in identifying the target channel, with the minimization carried out over all possible adaptive protocols. The structure of the most general adaptive protocol can be considered to be a quantum comb \cite{laurenza_channel_2018,chiribella_quantum_2008}.

\begin{figure}[ptb]
\centering
\includegraphics[width=0.7\linewidth]{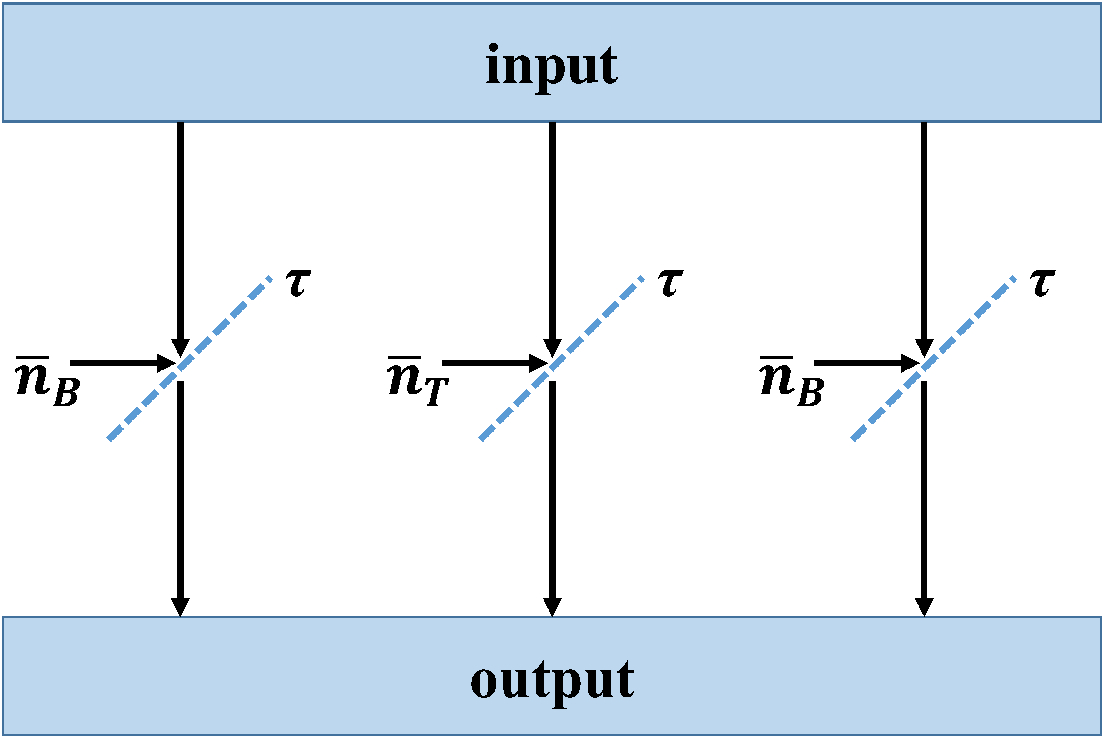}\caption{An example of the setup in the thermal loss case. Each thermal loss channel can be represented by a beamsplitter that mixes the input mode with an environmental thermal state. Thermal loss channels are parametrized by the transmissivity of the beamsplitter and the average photon number, $\bar{n}$, of the thermal state. We consider a sequence of thermal loss channels for which the beamsplitters all have the same transmissivity, $\tau$. One of the channels has a thermal state with a different average number of photons from the others; this is the target channel. The average number of photons in the thermal state of the target channel is denoted $\bar{n}_T$, whilst the average number of photons in the thermal state of the background channel is denoted $\bar{n}_B$. The task is to locate the target channel; in the case of this setup, it is the middle channel.}
\label{fig:setup}
\end{figure}

A schematic of a possible setup is given in Fig.~\ref{fig:setup}, which shows a sequence of three thermal loss channels with the same transmissivity, $\tau$. Two of these channels are background channels (with environmental noise $\bar{n}_B$) and one of the channels is the target channel (with environmental noise $\bar{n}_T$). At each channel use, we are allowed to send an input state through the sequence of channels, and this input state may be dependent on the previous channel outputs. Each channel is represented by a beamsplitter interaction with a thermal mode, and all of the beamsplitters have the same transmissivity, but the thermal mode with which the input modes interact is different for the target and background channels.

Any pair of one-mode, phase-insensitive, Gaussian channels with the same transmissivity is jointly teleportation covariant, using the Braunstein-Kimble (BK) protocol \cite{braunstein_teleportation_1998}. This means that both channels can be simulated using the same teleportation protocol, but with different resource states. In fact, using the BK protocol, a valid resource state for channel simulation is the asymptotic Choi matrix of the channel~\cite{Choi,Jamio,ChoiCV}. The Choi matrix of a channel is the output state when part of a maximally entangled state is passed through the channel. For bosonic systems, the maximally entangled state $\Phi$ is the limit for infinite squeezing of a sequence of two-mode squeezed vacuum (TMSV) states~\cite{weedbrook2012gaussian} $\Phi^{a}$, i.e., $\Phi=\lim_{a} \Phi^{a}$, where $a$ is the level of squeezing and each $\Phi^{a}$ has covariance matrix (CM)
\begin{align}
V_{\mathrm{in}}^a=\begin{pmatrix}
a \mathbb{I} &\sqrt{a^2-\frac{1}{4}}\mathbb{Z}\\
\sqrt{a^2-\frac{1}{4}}\mathbb{Z} &a \mathbb{I}
\end{pmatrix}.\label{eq:TMSV CM}
\end{align}
Therefore, the Choi matrix $\sigma_{\mathcal{E}}$ of a bosonic channel $\mathcal{E}$ is defined as the infinite-squeezing limit of a sequence of states $\{ \sigma^{a}_{\mathcal{E}} \}$ where the generic element is given by a TMSV state partially propagated through the channel, i.e., $\sigma^{a}_{\mathcal{E}}:=\mathcal{I}\otimes \mathcal{E}(\Phi^{a})$. In the following, when we work with an asymptotic Choi matrix $\sigma_{\mathcal{E}}$ we implicitly mean that this is the limit of an underlying `Choi sequence' $\{ \sigma^{a}_{\mathcal{E}} \}$. Correspondingly, the teleportation simulation over $\sigma_{\mathcal{E}}$ is meant to be an asymptotic operation, where the simulation is defined over the Choi sequence $\{ \sigma^{a}_{\mathcal{E}} \}$ after which the limit for infinite squeezing is taken~\cite{pirandola_fundamental_2017}. Note that Gaussian states, which all elements of the sequence are, are completely described by their CM and their first moments vector. For states in the Choi sequence, all elements of the first moments vector are 0.

The problem of CPF can be reduced to state discrimination between the $m$ possible outputs of the adaptive protocol used (with each outcome corresponding to a different target channel position). By bounding the fidelity between the different output states, we can find both upper and lower bounds for the minimum error probability $p_{\mathrm{err}}$ (optimized over all adaptive protocols) of state discrimination. The lower bound on the discrimination error between a sequence of $m$ states $\{\rho_i\}$, with probabilities $\{p_i\}$, is~\cite{montanaro2008lower}
\begin{align}
p_{\mathrm{err}}\geq \sum_{i>j}^m p_i p_j F^2(\rho_i,\rho_j),\label{eq:lower bound 0}
\end{align}
and the upper bound, based on the pretty good measurement (PGM) is~\cite{Barnum}
\begin{align}
p_{\mathrm{err}}\leq 2\sum_{i>j}^m \sqrt{p_i p_j} F(\rho_i,\rho_j),\label{eq:upper bound 0}
\end{align}
where $F$ is the Bures fidelity, defined as
\begin{align}
F(\rho_i,\rho_j)=\Tr \sqrt{\sqrt{\rho_i}\rho_j\sqrt{\rho_i}}.
\end{align}

\begin{figure}[ptb]
\centering
\includegraphics[width=1\linewidth]{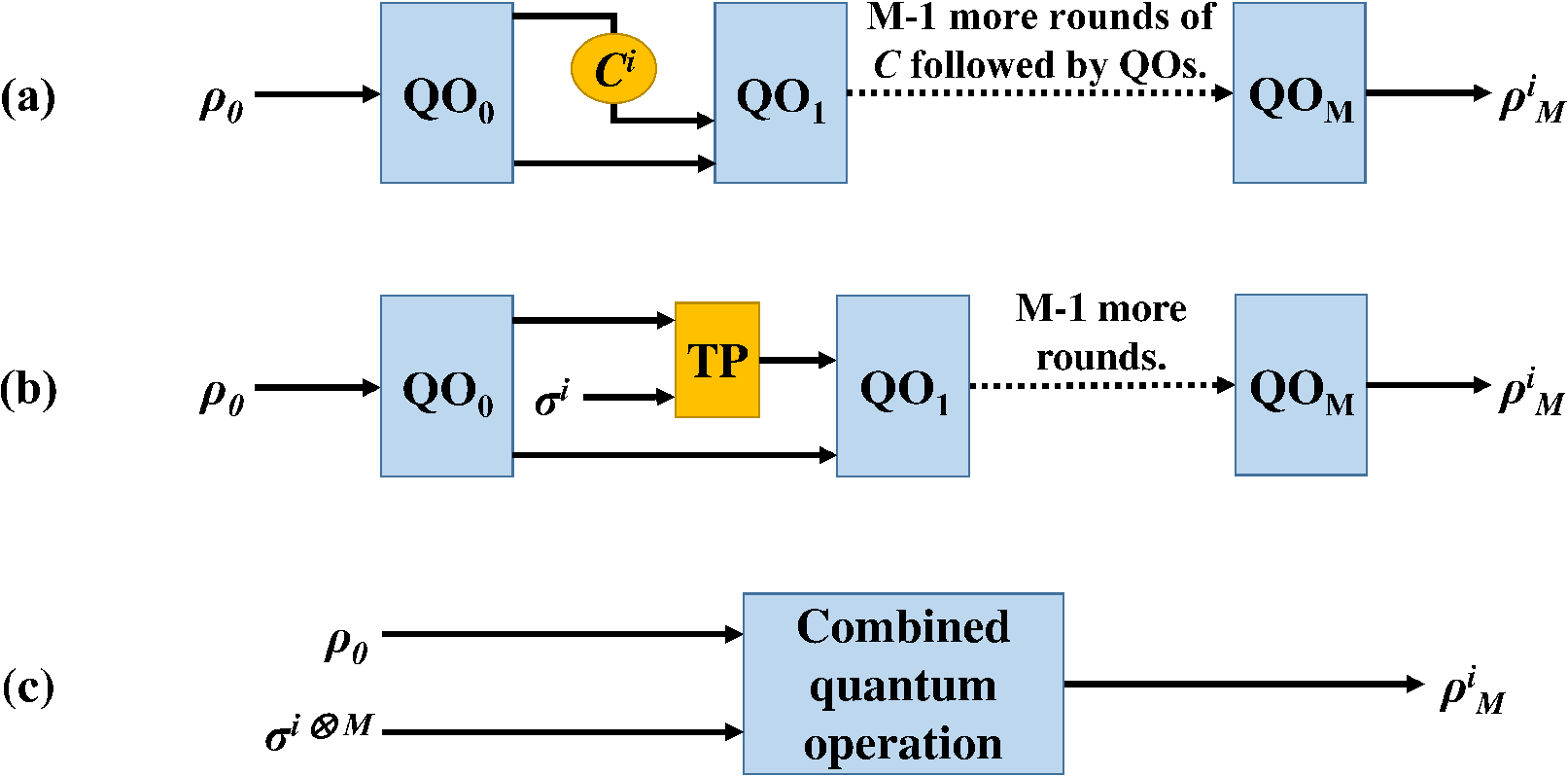}\caption{The reduction of a general adaptive discrimination protocol to a single round of quantum operations on a resource state. In panel (a), we have the most general discrimination protocol using $M$ uses of the sequence of channels. $\rho_0$ is some initial quantum state. We then apply some sequence of quantum operations (denoted by QO) interspersed with uses of the sequence of channels (denoted by $C^i$, where the label $i$ depends on the channel position). At each channel use, we may send a one-mode state through each of the channels in the sequence (and these modes are generally correlated with auxiliary modes that do not pass through the channels). Each round of quantum operations is allowed to be adaptive. This means that (i) entanglement can be present between ancillary modes of different quantum operations and (ii) measurements can be done on some subset of the modes and used to optimize following quantum operations. These measurements can always be delayed to the end of the protocol, by using controlled operations, so as to make all the QOs trace preserving. The final output of the adaptive protocol is denoted $\rho_0^i$; there are $m$ possible outputs depending on the channel position. Channel discrimination is then the task of discriminating between these $m$ different possible outputs, by means of an optimal collective quantum measurement (which may include all the delayed measurements). In panel (b), we simulate the channel with teleportation, using some teleportation protocol (TP) and a resource state ($\sigma^i$). Note that $\sigma^i$ is the resource state for the entire sequence of channels and is the tensor product of the resource states for teleportation of the $m-1$ background channels and the target channel, with the order of the subsystems determined by the label $i$. Note that neither the teleportation protocol nor the quantum operations depend on the label $i$ and so the entire discrimination protocol can be represented as some single fixed quantum operation on $\rho_0$ and $M$ copies of the resource state, $\sigma^i$. This representation is shown in panel (c).}
\label{fig:stretching}
\end{figure}

Since we can use the same teleportation protocol for both the target and the background channels, the entire discrimination protocol can be reduced, via stretching~\cite{pirandola_fundamental_2017,pirandola2017ultimate,pirandola2019fundamental}, to a single processor applied to different resource states (with the resource state depending on the position of the target channel). This adaptive-to-block reduction is shown in Fig.~\ref{fig:stretching}.

Since no trace preserving quantum operation can increase the distance between two quantum states (the fidelity of any two input states will be less than or equal to the fidelity of the resulting output states), the fidelity between the possible output states is lower bounded by the fidelity between the possible resource states. Let $\sigma^i_{M}$ be the resource state composed of $M(m-1)$ copies of the asymptotic Choi matrix of the background channel, $\sigma_{B}$, and $M$ copies of the asymptotic Choi matrix of the target channel, $\sigma_{T}$, arranged such that the $M$ copies of the asymptotic Choi matrix of the target channel is the $i$-th $2M$-mode subsystem. Note that each asymptotic Choi matrix consists of two modes. We can write
\begin{align}
\sigma^i_{M}=P_{1i}\left[\sigma_{T}^{\otimes M}\otimes\sigma_{B}^{\otimes M(m-1)}\right],
\end{align}
where the operator $P_{1i}$ swaps the first $2M$-mode subsystem with the $i$-th $2M$-mode subsystem. We can then lower bound the fidelity of any pair of output states of a discrimination protocol with $M$ channel uses using
\begin{align}
F(\rho^i_M,\rho^j_M)\geq F(\sigma^i_{M},\sigma^j_{M}).
\end{align}
Using the fact that each asymptotic Choi matrix in the resource is independent (i.e. using the tensor product structure of the resource states), we can write
\begin{align}
F(\sigma^i_{M},\sigma^j_{M})=F^{2M}(\sigma_{T},\sigma_{B}),
\end{align}
for all $i \neq j$.

More precisely, since the asymptotic Choi matrices, $\sigma_T$ and $\sigma_B$, are defined by the infinite-squeezing limit of two sequences of output states, $\{ \sigma_T^{a} \}$ and $\{ \sigma_B^{a} \}$, the fidelity functional is computed over the elements of the sequences and then the limit is taken, i.e., $F(\sigma_{T},\sigma_{B}):=\lim_a F(\sigma_{T}^{a},\sigma_{B}^{a})$. Then, it is important to notice that the bound $F(\rho^i_M,\rho^j_M) \geq F^{2M}(\sigma_{T},\sigma_{B})$ holds for any generally adaptive protocol $\mathcal{P}$. Therefore, we may write
\begin{equation}
    F_{i,j}:=\inf_{\mathcal{P}} F(\rho^i_M,\rho^j_M) \geq F^{2M}(\sigma_{T},\sigma_{B}).\label{onepart}
\end{equation}
At the same time, we note that this lower bound is achievable by a block protocol $\mathcal{P}_{\mathrm{block}}^{a}$  where $m$ copies of the
tensor product state $\Phi^{a \otimes M}$ are prepared and each TMSV state $\Phi^{a}$ is used for the single-probing of $\mathcal{I} \otimes \mathcal{E}_{B/T}$, so that the quasi-Choi matrix $\sigma_{B/T}^{a}$ is generated at the output for measurement. It is easy to see that, in the limit of infinite squeezing $a \rightarrow \infty$, this protocol achieves the performance at the right hand side of Eq.~(\ref{onepart}), so that we may write 
\begin{equation}
    F_{i,j}=F^{2M}(\sigma_{T},\sigma_{B}),~~\mathrm{for~any}~i,j.\label{secondpart}
\end{equation}

Let us optimize the error probability over all possible (generally adaptive) protocols $\mathcal{P}$. We define this optimal error probability as
\begin{align}
p_{\mathrm{err}}^{\mathrm{opt}}=\inf_{\mathcal{P}} p_{\mathrm{err}};
\end{align}
it is the smallest achievable error probability for any discrimination protocol. As a consequence of the reasoning above, and the inequalities in Eqs.~(\ref{eq:lower bound 0}) and (\ref{eq:upper bound 0}), we can write
\begin{align}
&p_{\mathrm{err}}^{\mathrm{opt}}\geq \sum_{i>j}^m p_i p_j F^{4M}(\sigma_{T},\sigma_{B}),\label{eq:lower bound 1}\\
&p_{\mathrm{err}}^{\mathrm{opt}}\leq 2\sum_{i>j}^m \sqrt{p_i p_j} F^{2M}(\sigma_{T},\sigma_{B}).\label{eq:upper bound 1}
\end{align}
Let us now assume that each channel position is equally likely, and so $p_i=\frac{1}{m}$ for every value of $i$. We can then carry out the sums in Eqs.~(\ref{eq:lower bound 1}) and (\ref{eq:upper bound 1}) and write
\begin{align}
&p_{\mathrm{err}}^{\mathrm{opt}}\geq \frac{m-1}{2m} F^{4M}(\sigma_{T},\sigma_{B}),\label{eq:lower bound 2}\\
&p_{\mathrm{err}}^{\mathrm{opt}}\leq (m-1)F^{2M}(\sigma_{T},\sigma_{B}).\label{eq:upper bound 2}
\end{align}

\subsection{Calculating the fidelity between Choi matrices}
\label{sec:fidelity}
We now must calculate the the fidelity between the (asymptotic) Choi matrices of the target and the background channels. A phase-insensitive, Gaussian channel~\cite{weedbrook2012gaussian} can be parametrized by two parameters: its transmissivity, $\tau$, and its induced noise, $\nu$. It transforms the CM of an input two-mode state, $V_{in}$, with the transformation
\begin{align}
V_{\mathrm{in}}\to\left(\mathbb{I}\oplus \sqrt{\tau}\mathbb{I} \right)V_{\mathrm{in}}\left(\mathbb{I}\oplus \sqrt{\tau}\mathbb{I} \right)^T+\left(0\oplus \nu\mathbb{I} \right),
\end{align}
where $\mathbb{I}$ is the 2 by 2 identity matrix. There are three main classes of phase-insensitive, Gaussian channels that we must consider: thermal loss channels, thermal amplifier channels and additive noise channels. Loss and amplifier channels both have $\nu\geq \frac{|1-\tau|}{2}$ (where we have chosen the shot noise to be $\frac{1}{2}$), but loss channels have $0\leq \tau <1$, whilst amplifier channels have $1<\tau$. Additive noise channels have $\nu\geq 0$ and $\tau=1$.

Passing the second mode of a TMSV state $\Phi^a$ with an average photon number per mode of $\bar{n}=a-\frac{1}{2}$ through a phase-insensitive, Gaussian channel results in the state with CM
\begin{align}
V_{\mathrm{out}}=\begin{pmatrix}
a \mathbb{I} &\sqrt{\tau\left(a^2-\frac{1}{4}\right)}\mathbb{Z}\\
\sqrt{\tau\left(a^2-\frac{1}{4}\right)}\mathbb{Z} &(a\tau +\nu) \mathbb{I}
\end{pmatrix},\label{eq:TMSVout}
\end{align}
where $\mathbb{Z}$ is the Z Pauli matrix.

The Bures fidelity of a pair of two-mode Gaussian states $\rho_i$ and $\rho_j$, with zero first moments and CM $V_i$ and $V_j$ is given by~\cite{marian2012uhlmann,banchi2015quantum}
\begin{align}
&F(\rho_i,\rho_j)=\frac{\sqrt{\chi}+\sqrt{\chi-1}}{\sqrt[4]{\det\left(V_i+V_j\right)}},\\
&\chi=2\sqrt{A}+2\sqrt{B}+\frac{1}{2},\\
&A=\frac{\det\left(\Omega V_i \Omega V_j -\frac{1}{4}\mathbb{I}\right)}{\det\left(V_i+V_j\right)},\\
&B=\frac{\det\left(V_i+\frac{i}{2}\Omega\right)\det\left(V_j+\frac{i}{2}\Omega\right)}{\det\left(V_i+V_j\right)},\\
&\Omega=\mathbb{I}\otimes\begin{pmatrix}
0 &1\\
-1 &0
\end{pmatrix}.
\end{align}
Using this expression, we can calculate the fidelity of a pair of output states of phase-insensitive, Gaussian channels (when the input state is a TMSV) with the same transmissivity.

In the case of thermal loss and amplifier channels, we define $\epsilon_T=\frac{\nu_{T}}{|1-\tau|}$ and $\epsilon_B=\frac{\nu_{B}}{|1-\tau|}$, where $\nu_T$ is the induced noise of the target channel, $\nu_B$ is the induced noise of the background channels, and $\tau$ is the transmissivity of all of the channels in the sequence. In fact, $\epsilon_T$ and $\epsilon_B$ give us the mean photon number of the environment for each channel, via the equation
\begin{align}
\bar{n}_{T(B)}=\epsilon_{T(B)}-\frac{1}{2}.
\end{align}

We find that the fidelity of the outputs of two such thermal loss or amplifier channels is analytically given by
\begin{align}
F_{\mathrm{loss/amp}}(\tau,\epsilon_T,\epsilon_B,a)=\frac{\sqrt{2}\left(\sqrt{\alpha+\beta}+\sqrt{\alpha-\beta}\right)}{\beta},\label{eq:fid_thermal}
\end{align}
where we define
\begin{align}
\begin{split}
\alpha=&\left(4\epsilon_T\epsilon_B+4a^2(4\epsilon_T\epsilon_B+1)\right.\\
&\left.+(4a^2-1)\sqrt{(4\epsilon_T^2-1)(4\epsilon_B^2-1)}\right)|1-\tau|^2\\
&+8a(\epsilon_T+\epsilon_B)\tau |1-\tau|+(1+\tau)^2,
\end{split}\\
\beta=&4\left(\tau+2a(\epsilon_T+\epsilon_B)|1-\tau|\right).
\end{align}

Taking the limit of this expression as $a\to\infty$, in order to obtain the fidelity between the Choi matrices, we get
\begin{align}
F_{\mathrm{loss/amp}}^{\infty}(\epsilon_T,\epsilon_B)=\frac{\sqrt{4\epsilon_T\epsilon_B+1+\sqrt{(4\epsilon_T^2-1)(4\epsilon_B^2-1)}}}{\sqrt{2}(\epsilon_T+\epsilon_B)}.\label{eq:choi_fid_thermal}
\end{align}
Note that we no longer have any explicit dependence on $\tau$.

Thus, our discrimination bounds for thermal loss or amplifier channels become
\begin{align}
p_{\mathrm{err}}^{\mathrm{opt}}\geq \frac{m-1}{2m} (F_{\mathrm{loss/amp}}^{\infty}(\epsilon_T,\epsilon_B))^{4M},\label{eq:lower bound thermal}\\
p_{\mathrm{err}}^{\mathrm{opt}}\leq (m-1)(F_{\mathrm{loss/amp}}^{\infty}(\epsilon_T,\epsilon_B))^{2M}.\label{eq:upper bound thermal}
\end{align}
The latter upper bound might become too large in some cases. Note that the error probability in randomly guessing the position of the target channel is equal to $(m-1)/m$. Combining this with the upper bound in Eq.~(\ref{eq:upper bound thermal}) leads to
\begin{equation}
p_{\mathrm{err}}^{\mathrm{opt}}\leq (m-1) \min\{m^{-1},(F_{\mathrm{loss/amp}}^{\infty}(\epsilon_T,\epsilon_B))\}.\label{eqForCap}
\end{equation}

In order to investigate the behaviour of $F_{\mathrm{loss/amp}}^{\infty}$, we re-parametrize Eq.~(\ref{eq:choi_fid_thermal}) in terms of the mean of $\epsilon_T$ and $\epsilon_B$, i.e.,
\begin{equation}
\epsilon_{\mathrm{av}}=\frac{\epsilon_T+\epsilon_B}{2},\label{meanEPS}
\end{equation}
and the absolute value of their difference, i.e.,
\begin{equation}
\epsilon_{\mathrm{dif}}=|\epsilon_T-\epsilon_B|.\label{diffEPS}
\end{equation}
Differentiating with regard to $\epsilon_{\mathrm{dif}}$, we get a negative semi-definite function and differentiating with regard to $\epsilon_{\mathrm{av}}$, we get a positive semi-definite function. This means that either increasing the difference in the average number of photons between the target and background channels (whilst keeping the mean fixed) or decreasing the mean of the $\epsilon$-values, whilst keeping the difference fixed, will decrease the minimum fidelity of the output states.

We now consider the case of additive noise channels. We find that the fidelity of the outputs of two such channels becomes
\begin{align}
F_{\mathrm{add}}(\nu_T,\nu_B,a)=\frac{2a\sqrt{\nu_T\nu_B}+\sqrt{(2a\nu_T+1)(2a\nu_B+1)}}{(2a(\nu_T+\nu_B)+1)}.\label{eq:fid_additive}
\end{align}
Taking the limit of this expression as $a\to\infty$, we get
\begin{align}
F_{\mathrm{add}}^{\infty}(\nu_T,\nu_B)=\frac{2\sqrt{\nu_T\nu_B}}{\nu_T+\nu_B}.\label{eq:choi_fid_additive}
\end{align}
We can again substitute this expression into Eqs.~(\ref{eq:lower bound 2}) and~(\ref{eq:upper bound 2}). Our discrimination bounds for additive noise channels become
\begin{align}
p_{\mathrm{err}}^{\mathrm{opt}}\geq \frac{m-1}{2m} (F_{\mathrm{add}}^{\infty}(\nu_T,\nu_B))^{4M},\label{eq:lower bound additive}\\
p_{\mathrm{err}}^{\mathrm{opt}}\leq (m-1)(F_{\mathrm{add}}^{\infty}(\nu_T,\nu_B))^{2M}.\label{eq:upper bound additive}
\end{align}

We now investigate the behaviour of $F_{\mathrm{add}}^{\infty}$ by re-parametrizing Eq.~(\ref{eq:choi_fid_additive}) in terms of $\nu_{\mathrm{av}}$ and $\nu_{\mathrm{dif}}$, where $\nu_{\mathrm{av}}$ is the mean of $\nu_T$ and $\nu_B$ and $\nu_{\mathrm{dif}}$ is the absolute value of the difference between them. Note that $\nu_{\mathrm{dif}}\leq 2\nu_{\mathrm{av}}$. We can then rewrite  Eq.~(\ref{eq:choi_fid_additive}) as
\begin{equation}
F_{\mathrm{add}}^{\infty}(r)=\sqrt{1-\frac{r^2}{4}},~~r=\frac{\nu_{\mathrm{dif}}}{\nu_{\mathrm{av}}}.
\end{equation}
Thus, we can see that the fidelity between the Choi matrices of two additive noise channels depends only on the ratio of $\nu_{\mathrm{dif}}$ to $\nu_{\mathrm{av}}$. Differentiating with regard to $r$, we see that the fidelity decays as $r$ increases.

\subsection{Classical limits}
\label{sec:classical limits}
Let us define a classical protocol as a protocol that restricts the states sent through the sequence of channels to an arbitrary mixture of coherent states. Since the Gaussian channels we are considering are phase-insensitive and since both the target and the background channels have the same transmissivity, enacting a phase-shift or displacement on the input states sent through the channels cannot affect the fidelity of the output states (since these unitary operations commute with the channels). The joint concavity of the Bures fidelity and the linearity of the channels means that the optimal classical input state (to minimize the fidelity between output states) is a single coherent state (not a mixture). As a result, the classical discrimination protocol that minimizes the lower bound on the error probability sends vacuum states through the channel at each channel use. This means that such protocols use only the passive signature of the channels.

We can obtain expressions for the minimum fidelity between output states for classical protocols by using our expressions for the fidelity between the output states using TMSV inputs in Eqs.~(\ref{eq:fid_thermal}) and (\ref{eq:fid_additive}) and setting $a=\frac{1}{2}$. This gives us the fidelity between the output states of the channels when the input state is a vacuum state.

In the case of thermal loss and amplifier channels, the minimum classical fidelity between output states is
\begin{align}
F_{\mathrm{loss/amp}}^{\mathrm{class}}(\tau,\epsilon_T,\epsilon_B)=\frac{\sqrt{\gamma+\delta}+\sqrt{\gamma-\delta}}{\delta},\label{eq:fid_class_thermal}
\end{align}
where we define
\begin{align}
&\gamma=4\epsilon_T\epsilon_B |1-\tau|^2+2(\epsilon_T+\epsilon_T)\tau|1-\tau|+(1+\tau^2),\\
&\delta=2\left(\tau+(\epsilon_T+\epsilon_T)|1-\tau|\right).
\end{align}
In the case of additive noise channels, the minimum classical fidelity between output states is
\begin{align}
F_{\mathrm{add}}^{\mathrm{class}}(\nu_T,\nu_B)=\frac{1}{\sqrt{(\nu_T+1)(\nu_B+1)}-\sqrt{\nu_T \nu_B}}.\label{eq:fid_class_additive}
\end{align}

We can now give upper and lower bounds on the error of classical discrimination protocols. We write
\begin{align}
p_{\mathrm{err}}^{\mathrm{class}}\geq \frac{m-1}{2m} (F^{\mathrm{class}})^{4M},\\
p_{\mathrm{err}}^{\mathrm{class}}\leq (m-1)(F^{\mathrm{class}})^{2M},\label{eq:classical_lower}
\end{align}
where the fidelity function is given in either Eq.~(\ref{eq:fid_class_thermal}) or Eq.~(\ref{eq:fid_class_additive}), depending on the class of channel.

\subsection{Quantum advantage}
\label{sec:quantum advantage}

We say that there is a quantum advantage if we can show that there exists some quantum discrimination protocol that gives a lower probability of error than any classical protocol. In order to prove a quantum advantage for channel position finding, we need to show that the lower bound on the error of classical protocols is larger than the upper bound on the error of all protocols. In other words, we must show that
\begin{align}
\frac{m-1}{2m} (F^{\mathrm{class}})^{4M}\geq (m-1)(F^{\infty})^{2M}.
\end{align}
This is equivalent to showing
\begin{align}
2M\ln\left(\frac{(F^{\mathrm{class}})^2}{F^{\infty}}\right)\geq \ln(2m).\label{eq:fid bound adv cond}
\end{align}
Noting that $\ln(2m)>0$, since $m\geq2$, we can see that the condition in Eq.~(\ref{eq:fid bound adv cond}) will always be met for sufficiently large $M$ (number of probes) as long as the condition
\begin{align}
(F^{\mathrm{class}})^2>F^{\infty}\label{eq:quant_adv_cond}
\end{align}
holds. Whether this condition is met depends only on the parameters of the target and background channels. Note that even if this condition is not met, it does not mean there is no quantum advantage; it could be the case that the bounds are not tight. In fact, in Section~\ref{sec: bounds protocols} we provide alternative bounds which can potentially show quantum advantage even in cases in which the condition in Eq.~(\ref{eq:quant_adv_cond}) is not met.

Unlike $F_{\mathrm{loss/amp}}^{\infty}$, the fidelity $F_{\mathrm{loss/amp}}^{\mathrm{class}}$ depends on the transmissivity $\tau$. In fact, differentiating, we find that $\frac{dF}{d\tau}\geq 0$ for $0\leq\tau<1$ and that $\frac{dF}{d\tau}\leq 0$ for $\tau>1$. Further, as $\tau\to 0$, we have $F_{\mathrm{loss/amp}}^{\mathrm{class}}\to F_{\mathrm{loss/amp}}^{\infty}$. This can be intuitively understood, since the entire channel discrimination process, including the coupling of the signal mode with the environment, can be regarded as a (generalized) measurement on the environmental modes. Thus, no matter how much entanglement the interacting modes have, the possible output states that the final measurement distinguishes between cannot have a lower (pairwise) fidelity than the possible configurations of environmental modes that are being discriminated between. In other words, the infinite squeezing case is equivalent to a direct measurement on the environmental modes before they are mixed with the signal states, whilst, in any finite energy scenario, we send signal states to interact with the environmental modes and then measure the signal states. Since the $\tau=0$ case corresponds to the signal states being completely replaced by the environmental modes, the classical protocol, in this case, is also a direct measurement on the environmental modes. Consequently, in the case of thermal loss channels, for all values of $\epsilon_T$ and $\epsilon_B$, there is some threshold value of $\tau$ such that channels with $\tau$ below the threshold do not meet the condition in Eq.~(\ref{eq:quant_adv_cond}). Setting $\tau=\frac{1}{2}$, we find that$\frac{(F^{\mathrm{class}})^2}{F^{\infty}}\leq 1$, and hence the inequality in Eq.~(\ref{eq:quant_adv_cond}) does not hold for any channel ensemble with $\tau\leq \frac{1}{2}$. For further details, see Appendix~\ref{appendix:fidelity}.

\begin{figure}[t]
\centering
\includegraphics[width=0.9\linewidth]{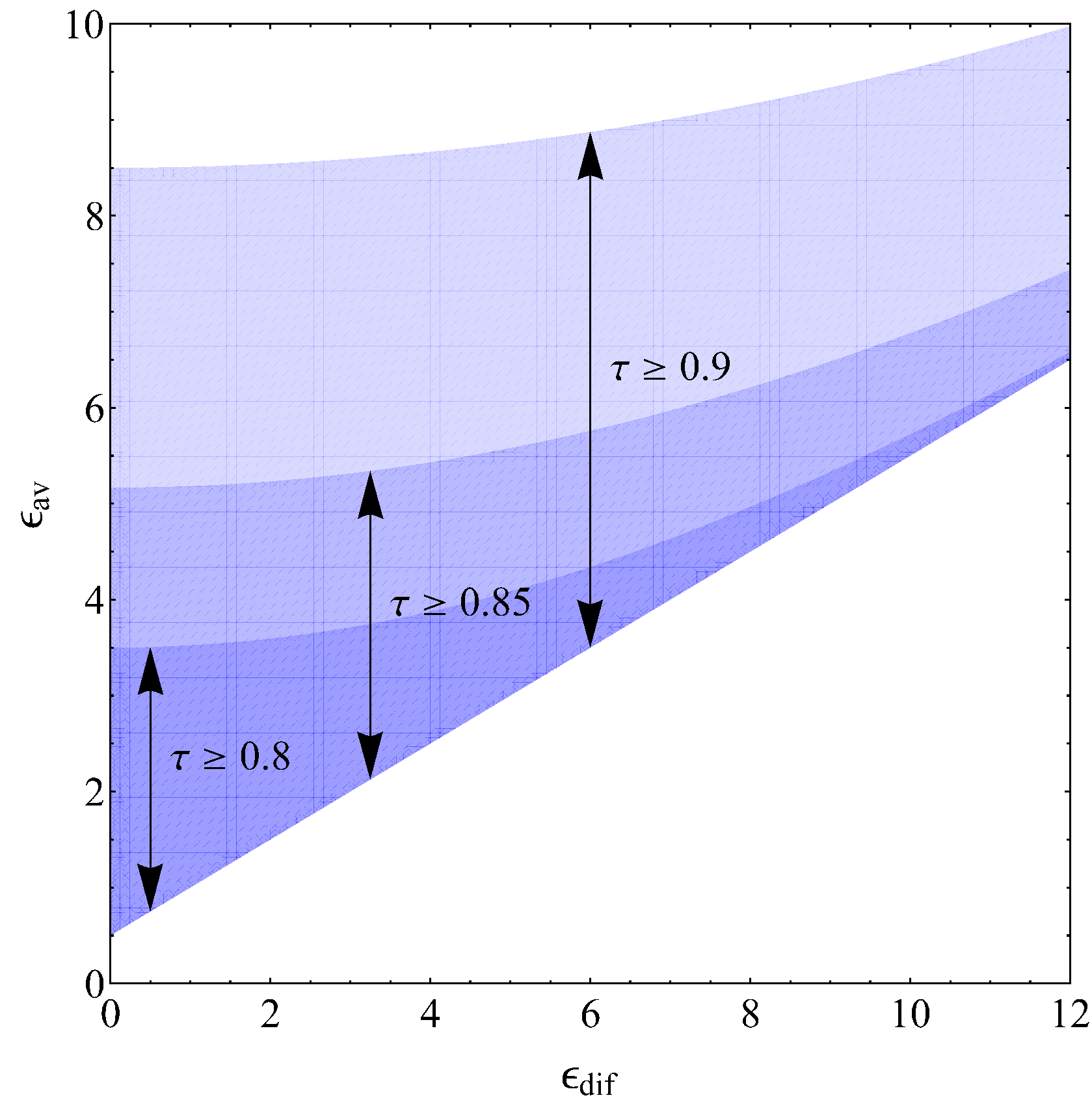}\caption{Regions in which we can prove a quantum advantage for thermal loss channels, as a function of their noise difference $\epsilon_{\mathrm{dif}}$ and mean noise $\epsilon_{\mathrm{av}}$, for different values of the transmissivity $\tau$. Note that the region for a higher value of $\tau$ completely contains the region for any lower value of $\tau$. The minimum value of $\epsilon_{\mathrm{av}}$ for fixed $\epsilon_{\mathrm{dif}}$ is $\frac{\epsilon_{\mathrm{dif}}+1}{2}$, since neither $\epsilon_T$ nor $\epsilon_B$ can be less than $\frac{1}{2}$.}
\label{fig:quantum_adv}
\end{figure}

Fig.~\ref{fig:quantum_adv} illustrates the region in which we meet the condition in Eq.~(\ref{eq:quant_adv_cond}) (and so can prove a quantum advantage for some number of probes), in the case of thermal loss channels, for a few choices of transmissivity, $\tau$. The plot is in terms of $\epsilon_{\mathrm{dif}}$ and $\epsilon_{\mathrm{av}}$, as defined in Eqs.~(\ref{meanEPS})-(\ref{diffEPS}). We see that higher transmissivities result in a larger region in which we can prove a quantum advantage. Further, as $\epsilon_{\mathrm{dif}}$ increases, the region in which we can prove quantum advantage narrows (in terms of the allowed values of $\epsilon_{\mathrm{av}}$).

The condition for the inequality in Eq.~(\ref{eq:quant_adv_cond}) to hold takes a simple form for additive noise channels. We again re-parametrize in terms of $\nu_{\mathrm{av}}$ and $\nu_{\mathrm{dif}}$. We can then write the condition purely in terms of $\nu_{\mathrm{av}}$. Thus, we find that for a sequence of additive noise channels, we will always have a quantum advantage for some number of probes as long as
\begin{align}
\nu_{\mathrm{dif}}>\frac{\sqrt{32\nu_{\mathrm{av}}^4-8\nu_{\mathrm{av}}^2-8\nu_{\mathrm{av}}-1-(4\nu_{\mathrm{av}}+1)\sqrt{8\nu_{\mathrm{av}}+1}}}{2\sqrt{2}\nu_{\mathrm{av}}}.
\end{align}

\subsection{Bounds from specific protocols}
\label{sec: bounds protocols}
We can consider specific discrimination protocols; these can provide benchmarks for both the classical (entanglement-free) and entangled cases. In the classical case, we have vacuum input. In this case, the return state is thermal, therefore a photon counting measurement coupled with the maximum-likelihood estimation (MLE) gives the Helstrom performance~\cite{Helstrom_1976}. In this protocol, we carry out photon counting on each of the return states, and simple derivation shows that the MLE decision rule reduces to choosing the channel with the maximum/minimum photon count, i.e., we estimate the target channel to be
\begin{align}
\arg\max_s N_s, \mbox{if }\bar{n}_T> \bar{n}_B,
\end{align}
and
\begin{align}
\arg\min_s N_s, \mbox{if }\bar{n}_T<\bar{n}_B,
\end{align}
where $s$ is an index labelling the channels in the sequence and $N_s$ denotes the total number of photons counted from the return states of channel $s$ (cumulatively, over all M channel uses). 

\begin{figure}[ptb]
\centering
\includegraphics[width=1\linewidth]{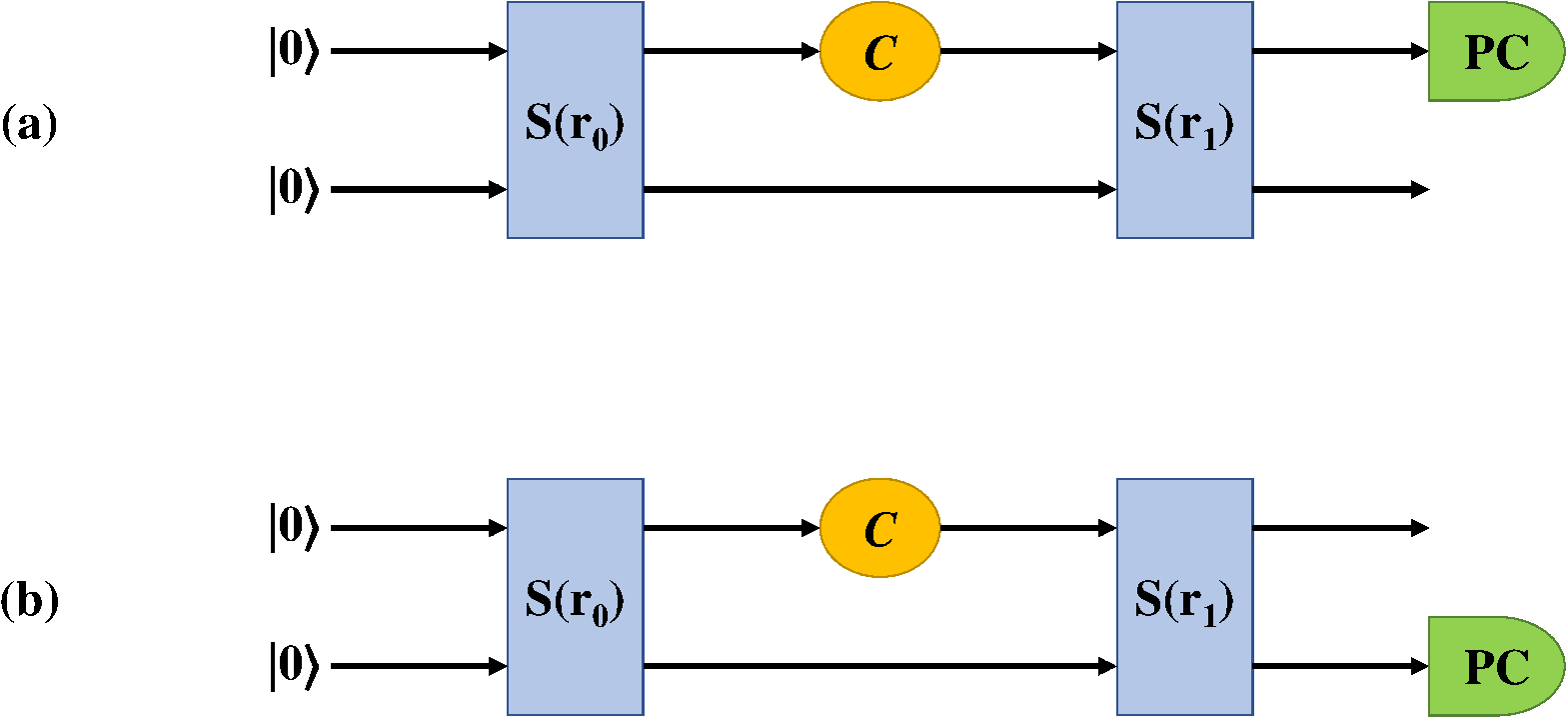}\caption{The setup for a CPF protocol that provides a benchmark for the general quantum case. In panel (a), we have the protocol for the thermal loss case and in panel (b), we have the protocol for the thermal amplifier case. In both cases, we begin by carrying out two-mode squeezing on a vacuum state, with squeezing parameter $r_0$, as given in Eq.~(\ref{eq:sq_init}). This is denoted $\mathrm{S(r_0)}$. We then pass one of the modes through the channel, denoted $C$, and then carry out two-mode squeezing again, this time with squeezing parameter $r_1$. Finally, we carry out a photon counting measurement (denoted PC) on one of the modes and trace over the other mode. This process is repeated $M$ times (where $M$ is the number of probes used) for every channel in the sequence. Note that in the thermal loss case, the measurement is carried out on the channel mode, whilst in the thermal amplifier case, the measurement is carried out on the idler mode.}
\label{fig:MLE_protocol}
\end{figure}

We can consider a similar protocol involving entanglement, in the cases of thermal loss and amplifier channels. In these cases, we can get thermal return states by sending TMSV states through the channels, carrying out anti-squeezing operations on the return states and then tracing over one of the two modes. For each probe sent through one of the channels, we start by carrying out two-mode squeezing on a pair of vacuum modes, with squeezing parameter
\begin{align}
r_0=\frac{1}{2}\ln\left(2a+\sqrt{4a^2-1}\right).\label{eq:sq_init}
\end{align}
This results in the TMSV state $\Phi^{a}$, which has an average photon number per mode of $\bar{n}=a-\frac{1}{2}$ and the CM given by Eq.~(\ref{eq:TMSV CM}). The first mode is kept as an idler, whilst the second mode is passed through the channel. Each individual channel output state will then have a CM of the form in Eq.~(\ref{eq:TMSVout}); we then carry out two-mode squeezing on the state, with squeezing parameter
\begin{align}
r_1=\frac{1}{2}\ln\left(\frac{|1-\sqrt{\tau|}}{1+\sqrt{\tau}}\right).\label{eq:sq_param}
\end{align}
For a thermal loss channel, we discard the idler mode; the resulting state has the CM
\begin{align}
V_{\mathrm{ret},\mathrm{loss}}^a&=\mathrm{Disc}_{1} \left[S(r_1)V_{\mathrm{out},\mathrm{loss}}^{a}S^T(r_1)\right]\\
&=\frac{\nu+2a\tau-\tau\sqrt{4a^2-1}}{|1-\tau|}\mathbb{I},\label{eq:V_ret}
\end{align}
where $S$ is the two-mode squeezing matrix, given by
\begin{align}
S(r)=\begin{pmatrix}
\cosh(r)\mathbb{I} &\sinh(r)\mathbb{Z}\\
\sinh(r)\mathbb{Z} &\cosh(r)\mathbb{I}
\end{pmatrix},
\end{align}
and where $\mathrm{Disc}_1$ indicates that we discard the first (idler) mode. We can get a return state with the same form for an amplifier channel by carrying out the same process, but tracing over the other mode (the mode which passed through the channel). In other words, we have
\begin{align}
V_{\mathrm{ret},\mathrm{amp}}^a&=\mathrm{Disc}_{2} \left[S(r_1)V_{\mathrm{out},\mathrm{amp}}^{a}S^T(r_1)\right]\\
&=\frac{\nu+2a\tau-\tau\sqrt{4a^2-1}}{|1-\tau|}\mathbb{I}.
\end{align}
This protocol is illustrated in Fig.~\ref{fig:MLE_protocol}.

We now note that the CM in Eq.~(\ref{eq:V_ret}) has finite energy, even in the limit of infinite squeezing ($a \rightarrow \infty$). Letting $V_{\mathrm{ret},T(B)}^{\infty}$ be the asymptotic return state from the target (background) channel (for either a thermal loss or a thermal amplifier channel), we find that
\begin{align}
V_{\mathrm{ret},T(B)}^{\infty}=\frac{\nu_{T(B)}}{|1-\tau|}\mathbb{I}=\epsilon_{T(B)}\mathbb{I}.
\end{align}
Hence, we can get thermal return states even in the case of infinite entanglement. Note that these are the same return states we would get in the classical case if the channels had a transmissivity of 0. Note too that we cannot enact this protocol in the additive noise case, since our expression in Eq.~(\ref{eq:sq_param}) for the squeezing parameter $r_1$ diverges as $\tau\to 1$. We can then carry out photon counting measurements on the return states and estimate the target channel using the MLE.

We now calculate the success probability of the MLE. The probability that a thermal mode with average photon number $\bar{n}$ is measured to have $k$ photons is given by
\begin{align}
P_{\bar{n}}(k)=\frac{\bar{n}^k}{(\bar{n}+1)^{k+1}}.
\end{align}
We then calculate the probability that $M$ thermal modes, with the same average photon number of $\bar{n}$, are measured to have a total of $k$ photons, by replacing the thermal distribution with a sum of independent and identically distributed (iid) thermal distributions. We find that this probability is given by
\begin{align}
P_{\bar{n},M}(k)={k+M-1\choose k}\left(\frac{\bar{n}}{1+\bar{n}}\right)^k \left(\frac{1}{1+\bar{n}}\right)^M,
\end{align}
where the binomial coefficient accounts for the different ways in which the photons can be distributed across the measured modes. From this we can calculate the probability that the $M$ modes are measured to have fewer than $n_c$ photons in total:
\begin{align}
{\rm pr}_{\bar{n},M}({\rm count}<n_c)=\sum_{k=0}^{n_c-1} P_{\bar{n},M}(k).
\end{align}

Let us first consider the case in which $\bar{n}_T>\bar{n}_B$. In this case the MLE gives the correct answer when all of the background channels have return states that are measured to have fewer photons than those of the target channel. We must also consider the possibility that the return states of one or more of the background channels are measured to have the same number of photons as the return states of the target channel (but not more). In this case, we choose randomly between the channels that gave the highest photon counts. This gives a total success probability (for the entangled case) of
\begin{align}
\begin{split}
p^{\mathrm{MLE}}_{\mathrm{succ},\bar{n}_T>\bar{n}_B}=&\sum_{c=1}^{m} \frac{1}{c} \sum_{n_c=0}^\infty \left[{\rm pr}_{\bar{n}_{B},M}({\rm count}<n_c)\right]^{m-c}\\
&\times P_{\bar{n}_{T},M}(n_c){m-1\choose c-1}(P_{\bar{n}_{B},M}(n_c))^{c-1}.
\end{split}
\end{align}
Here, the index $c$ is the number of channels with the same photon count (hence, $c=1$ is the case in which all of the background channels give a lower photon count than the target channel). The factor of $\frac{1}{c}$ comes from the random choice when multiple channels give the same photon count. Note that in the case of $n_c=0$, the only non-zero contribution is in the case $c=m$, corresponding to a photon count of 0 for the target and all of the background channels. If this occurs, there is a $\frac{1}{m}$ chance of the correct channel being randomly guessed to be the target channel. In this case, we define
\begin{align}
{\rm pr}_{\bar{n}_{B},M}({\rm count}<0)^{0}=1.
\end{align}

Extension to the case in which $\bar{n}_T<\bar{n}_B$ can be done trivially, by writing
\begin{align}
{\rm pr}_{\bar{n},M}({\rm count}>n_c)=1-{\rm pr}_{\bar{n},M}({\rm count}<n_c+1).
\end{align}
Then we have a success probability of
\begin{align}
\begin{split}
p^{\mathrm{MLE}}_{\mathrm{succ},\bar{n}_T<\bar{n}_B}=&\sum_{c=1}^{m} \frac{1}{c} \sum_{n_c=0}^\infty \left[{\rm pr}_{\bar{n}_{B},M}({\rm count}> n_c)\right]^{m-c}\\
&\times P_{\bar{n}_{T},M}(n_c){m-1\choose c-1}(P_{\bar{n}_{B},M}(n_c))^{c-1}.
\end{split}
\end{align}

In both cases, the error probability is given by
\begin{align}
p^{\mathrm{MLE}}_{\mathrm{err}}=1-p^{\mathrm{MLE}}_{\mathrm{succ}}.
\end{align}
Note that for the classical MLE error probabilities, we simply substitute $\bar{n}_{T(B)}$ with the average photon numbers of the classical return states, i.e. $\bar{n}_{T(B)}|1-\tau|$.

This quantity can be easily numerically calculated. Using this semi-analytic benchmark, we can show a quantum advantage with a lower value of $M$ than is required for the condition in Eq.~(\ref{eq:fid bound adv cond}) to be met. This is demonstrated in Fig.~\ref{fig:imaging}. It is also useful as it is based on a protocol that can be easily implemented.

The scaling of the MLE error with the number of subsystems is of interest. We can upper bound the error in the case of $m$ subsystems in terms of the success probability for 2 subsystems, which we will call $p^{\mathrm{MLE}}_{\mathrm{succ},2}$. The error probability for $m$ subsystems then obeys the inequality
\begin{align}
p^{\mathrm{MLE}}_{\mathrm{err},m}\leq 1-(p^{\mathrm{MLE}}_{\mathrm{succ},2})^{m-1}=1-\left(1-p^{\mathrm{MLE}}_{\mathrm{err},2}\right)^{m-1},\label{eq:MLE scaling}
\end{align}
since the target channel having a higher photon count than one background channel cannot decrease the probability that it will have a higher photon count than a different background channel. In fact, this bound is an overestimate for any $m>2$, since the conditional probability that the target channel has a higher photon count than one background channel, given that it has a higher photon count than a different background channel, is more than $p^{\mathrm{MLE}}_{\mathrm{succ},2}$. This can be understood by considering the iid outcomes of 3 (6-sided) dice rolls denoted $a$, $b$ and $c$. The probability that $a>b$ is the same as the probability that $a>c$ and is equal to $\frac{5}{12}$, however the probability that $a>c$ given that $a>b$ is more than $\frac{5}{12}$, since the condition makes it less likely that $a$ is a small number and more likely that $a$ is a large number. Expanding the inequality in Eq.~(\ref{eq:MLE scaling}) to the first order in $p^{\mathrm{MLE}}_{\mathrm{err},2}$, we get
\begin{align}
p^{\mathrm{MLE}}_{\mathrm{err},m}\leq (m-1)p^{\mathrm{MLE}}_{\mathrm{err},2}.\label{eq:MLE scaling UB}
\end{align}
This inequality is strict for $m>2$. This means that the MLE error scales more slowly with $m$ than the upper bound in Eq.~(\ref{eq:upper bound 2}), which is based on the PGM. However, for some sets of channel parameters, the upper bound in Eq.~(\ref{eq:MLE scaling UB}) can be close to the actual value of $p^{\mathrm{MLE}}_{\mathrm{err},m}$.

It is also of note that, whilst the bounds based on the fidelity are symmetric under the exchange of $\nu_T$ and $\nu_B$, the MLE bound is not (for more than two subsystems). Thus, using this protocol in one of our applications, we may achieve a different error probability for finding a single cold pixel in a hot background than for finding a single hot pixel in a cold background.

\subsection{Applications of the bounds}
\label{sec:applications}
Let us consider some physical applications of these bounds. One possible scenario in which one may need to discriminate between various channels with the same transmissivity is thermal imaging. The sequence of channels could represent a sequence of pixels that is being probed with microwave or infrared radiation, where we know that one pixel is hotter (or colder) than its surroundings and want to know its location. Alternatively, we could be imaging a surface with a microscope and want to find the frequency at which a source on the surface is emitting radiation. The different channels would then represent different frequencies. These tasks can both be modelled as a CPF task over a sequence of thermal loss channels with the same transmissivity.

\begin{figure}[ptb]
\centering
\includegraphics[width=0.9\linewidth]{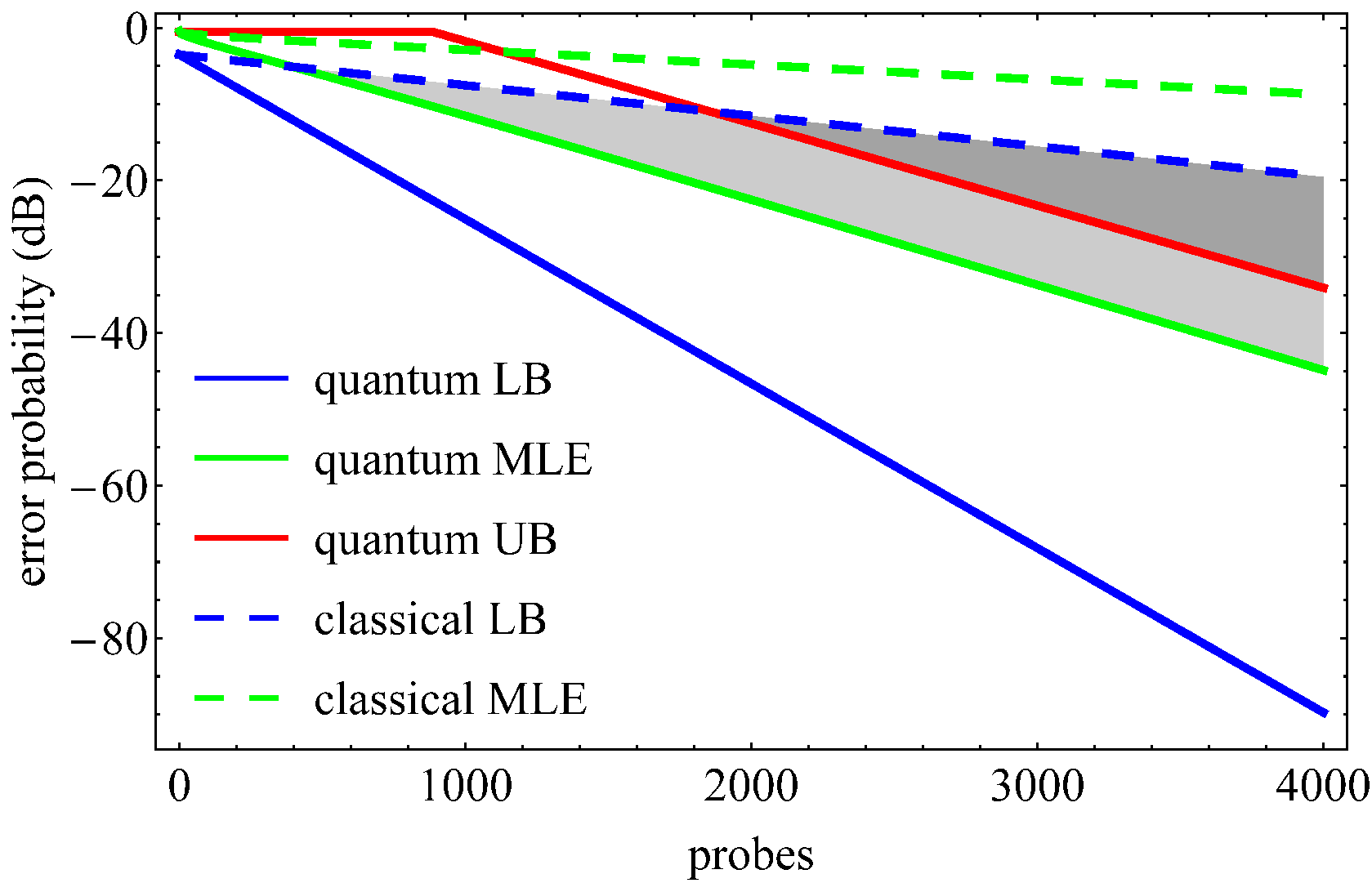}\caption{Error probability in decibels (dB), $10 \log_{10}(p_{\mathrm{err}})$, as a function of the number of the probes per pixel, for a thermal imaging task in which a sequence of $m=9$ pixels, each of area $4000~\mathrm{\mu m^2}$, is probed using microwaves (with wavelength 1~mm). The transmissivity of each pixel is 0.99 and the goal is finding the one pixel at temperature $247.56$~K ($-25.59$\textdegree{}C, $\epsilon_T=21$) in a background of pixels at temperature $272.76$~K ($-0.39$\textdegree{}C, $\epsilon_B=23.2$). Lower and upper bounds on the error probability are given for general quantum protocols (labelled ``quantum LB" and ``quantum UB") and a lower bound on the error is given for classical protocols (labelled ``classical LB"), for differing numbers of states sent through the channels (probes). Benchmarks based on the MLE are also shown for both the quantum and the classical cases (labelled ``quantum MLE" and ``classical MLE"). For the quantum upper bound, we use the expression in Eq.~(\ref{eqForCap}). For a large number of probes (in this case, greater than or equal to 1854), the upper bound on the error of quantum protocols is smaller than the lower bound on the error of classical protocols, proving we have a quantum advantage (in the darker shaded area). However, a much smaller number of probes (396) is required for the bound based on the MLE in the quantum case to beat the classical lower bound, and hence we are able to show a quantum advantage for any number of probes greater than 395 (in the lighter shaded area).}
\label{fig:imaging}
\end{figure}

In Fig.~\ref{fig:imaging}, we consider an imaging task, in which a colder pixel must be located from a sequence of $9$ pixels, each of which has an area, $A$, of $4000~\mathrm{\mu m^2}$. We consider a case in which imaging is carried out in the microwave range (with a wavelength of $1$~mm), with a background temperature of ${\sim}-0.39$\textdegree{}C, a target temperature of ${\sim}-25.59$\textdegree{}C, and high transmissivity. We assume that our detectors are very close to the pixels and that our imaging pulses have a time duration, $t$, of $100$~ns. We also assume that the pulses are transform-limited and so set the bandwidth of detection to $2.5$~MHz. This is in line with the fact that a transform-limited pulse has a time-bandwidth product (in terms of the variances) of $\frac{1}{4}$ \cite{siegman_1986}.

We find the mean photon numbers by calculating the induced noise, which is independent of the transmissivity. Planck's law states that the spectral radiance of a black body, at a frequency $f$, is given by
\begin{align}
R(f,T)=\frac{2hf^3}{c^2(e^{\frac{hf}{kT}}-1)},
\end{align}
where $c$ is the speed of light, $h$ is Planck's constant, $k$ is the Boltzmann constant, and $T$ is the temperature of the pixel. By dividing $R$ by $hf$, we obtain the number of photons emitted per unit time, per unit area of the pixel into an infinitesimal frequency range and into a unit solid angle. We must then integrate $\frac{R}{hf}$ over the bandwidth of the detector and multiply it by the duration of the imaging pulse, $t$, the solid angle over which the detector collects photons, $\Sigma$, and the area of the pixels, $A$, in order to obtain the induced noise, $\nu$. We therefore write
\begin{align}
\nu_{B/T}=A\Sigma t\int_{f_{\mathrm{min}}}^{f_{\mathrm{max}}} \frac{2f^2}{c^2(e^{\frac{hf}{kT_{B/T}}}-1)}df,
\end{align}
where $T_{B/T}$ is the temperature of the background/target pixel and $f_{\mathrm{min/max}}$ is the minimum/maximum frequency in our frequency range. We set $\Sigma=2\pi$ (i.e. we assume that the detector collects all light emitted in one hemisphere normal to the surface of the pixel). This is justified by our assumption that the detector is close to the pixels. If the detector were further away, we could adjust $\Sigma$ accordingly (and may have to reduce the transmissivity, $\tau$). Dividing $\nu_B$ and $\nu_T$ by $|1-\tau|$ gives the values of $\epsilon_B$ and $\epsilon_T$ respectively.

Note that, for the bounds based on fidelity, swapping $\epsilon_T$ and $\epsilon_B$ does not affect the calculations, so these would be the same if the task were to find a target pixel at temperature $-0.39$\textdegree{}C in a background of pixels at $-25.59$\textdegree{}C. This is not the case for the benchmark based on the MLE. From Fig.~\ref{fig:imaging}, we see that we can prove a quantum advantage for a large number of channel uses (probes). We also see that the (quantum) MLE bound enables us to show a quantum advantage at a much lower value of $M$ than the fidelity-based quantum upper bound.

Before considering the next example, it is also worth noting that it is likely that the classical lower bound (blue dashed) in Fig.~\ref{fig:imaging} is not tight, since we see a gap between it and the classical MLE performance (green dashed). Therefore quantum advantage is likely to hold for any number of probes, since we see that the quantum MLE (green solid) beats the classical MLE (green dashed) for any number of probes. A future study might be able to prove such a quantum advantage.

Another scenario in which one may wish to discriminate between thermal loss channels with different noises could arise in quantum communications. One may know that one of a sequence of communications lines has a higher excess noise than the others, perhaps due to the presence of an eavesdropper, and may wish to localize the eavesdropper by finding the channel with the higher excess noise.

\begin{figure}[t]
\centering
\includegraphics[width=0.9\linewidth]{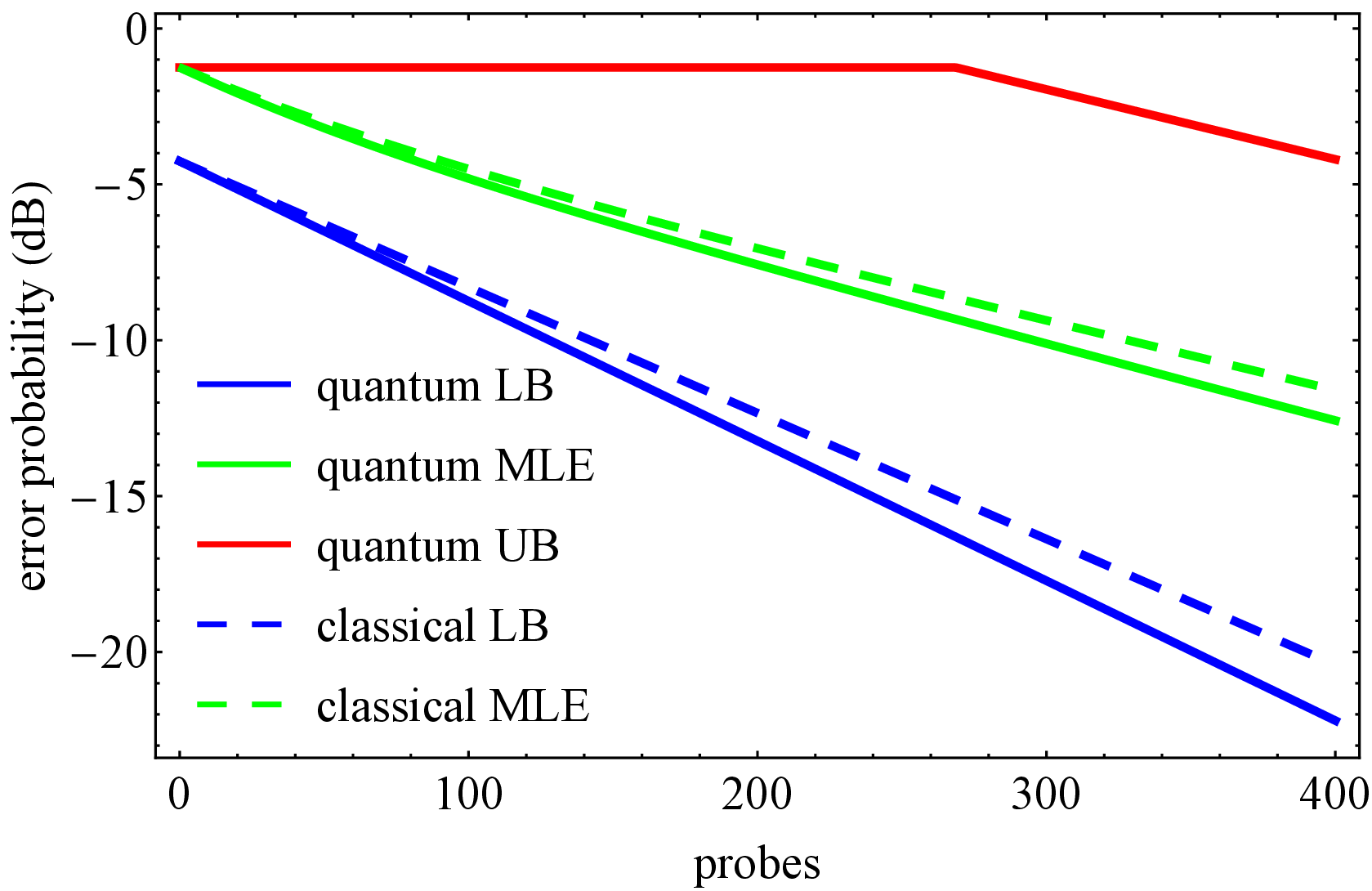}\caption{Error probability in decibels versus number of probes per communication line for the problem of eavesdropper localization. We consider a transmissivity of 0.1, corresponding to a loss of 10~dB. The background channels have an excess noise of 0.01, whilst the channel with the eavesdropper has an excess noise of 0.1. Lower and upper bounds on the error probability are given for general quantum protocols (labelled ``quantum LB" and ``quantum UB") and a lower bound on the error is given for classical protocols (labelled ``classical LB"). Benchmarks based on the MLE are shown for both the quantum and the classical cases (labelled ``quantum MLE" and ``classical MLE"). In this case, the quantum upper bound never goes below the classical upper bound, so we are not able to prove a quantum advantage.}
\label{fig:eavesdropper}
\end{figure}

\begin{figure}[t!]
\centering
\includegraphics[width=0.9\linewidth]{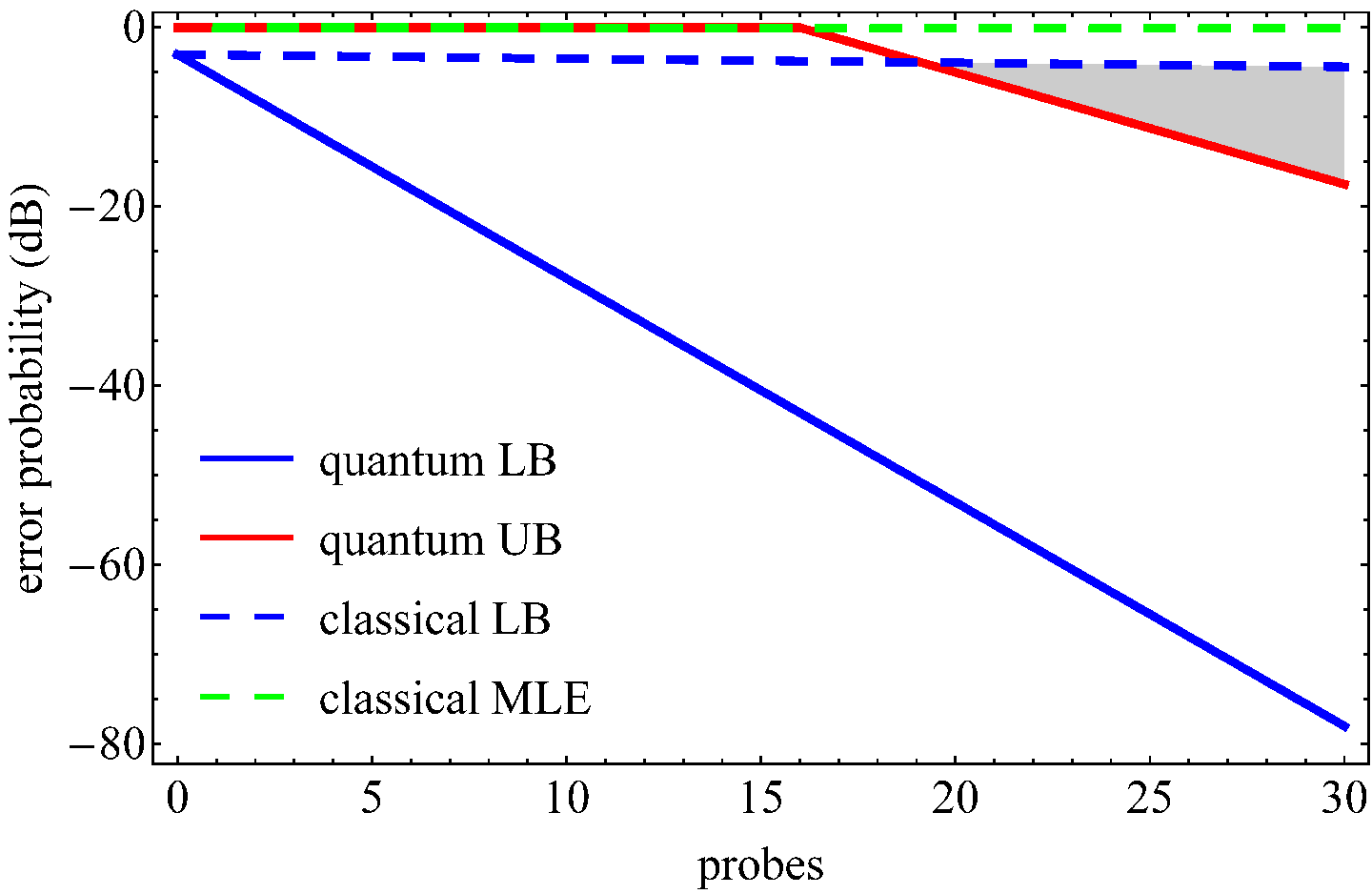}\caption{Error probability in decibels versus number of probes per channel for the problem of additive noise localization. We want to find the channel with the lower induced noise from a sequence of 100 additive-noise channels. The background channels have an induced noise of 0.03, whilst the target channel has an induced noise of 0.01. Lower and upper bounds on the error probability are given for general quantum protocols (labelled ``quantum LB" and ``quantum UB") and a lower bound on the error is given for classical protocols (labelled ``classical LB"). The benchmark based on the MLE is shown for the classical case (labelled ``classical MLE"). For a number of probes greater than or equal to 20, the upper bound on the error of quantum protocols is smaller than the lower bound on the error of classical protocols, proving we have a quantum advantage (in the shaded area).}
\label{fig:position}
\end{figure}

This scenario is illustrated in Fig.~\ref{fig:eavesdropper}, where we consider transmission over communication lines with a loss of 10~dB. Excess noise is expressed in dimensionless shot noise units and is defined in terms of the transmissivity and the thermal number of the channel as $\epsilon = \tau^{-1} (1-\tau) \bar{n}$~\cite{CryptoReview}.
We consider background excess noises of 0.01 and an excess noise for the eavesdropper of 0.1. In this case, we cannot prove a quantum advantage, although the quantum lower bound is lower than the classical lower bound. This is in accordance with the fact that we cannot meet the condition in Eq.~(\ref{eq:quant_adv_cond}) with any channel ensemble that has $\tau\leq\frac{1}{2}$. The quantum MLE benchmark is also lower than the classical MLE benchmark, but does not go below the classical lower bound. This is again likely to be caused by the classical lower bound not being tight.

Another possibility is that we could have a multi-mode cable with multiple frequency channels and wish to find a channel with lower noise than the others. This is another case of discrimination between a sequence of thermal loss channels with different noises. If the transmissivity is high enough (for instance, for a short-range cable) we could potentially also model this scenario as a sequence of additive noise channels.

Fig.~\ref{fig:position} illustrates this situation. We consider a sequence of 100 additive noise channels and want to find the channel with the lower induced noise. The background channels have an induced noise of 0.03 and the target channel has an induced noise of 0.01. We can show a quantum advantage for a number of probes greater than or equal to 20. Note that, whilst we can provide a classical benchmark based on the MLE, we cannot provide a quantum MLE benchmark in the additive noise case. This is due to the fact that the squeezing parameter in Eq.~(\ref{eq:sq_param}) diverges as $\tau\to 1$, meaning that the protocol shown in Fig.~\ref{fig:MLE_protocol} cannot be enacted in the additive noise case.

\section{Conclusion}
In this work we have considered the problem of channel-position finding with a passive signature, where the aim is to localize a target channel in a sequence of background channels with the same transmissivity/gain but a different induced noise. The problem can therefore be seen as a problem of environment localization. We have considered this model in the setting of bosonic systems, considering such a localization with phase-insensitive Gaussian channels, such as thermal-loss channels (with the same transmissivity but different thermal noise), noisy quantum amplifiers (with the same gain but different thermal noise), and additive noise channels (with different added noise).

Using channel simulation and protocol stretching, we have determined upper and lower bounds for the optimal error probability for environment localization. These bounds hold for the most general, adaptive, multi-ary quantum discrimination protocols. By comparison with a classical benchmark, associated with the optimal performance achievable by coherent states, we have determined the mathematical conditions to prove a quantum advantage. If these conditions on the noise parameters are satisfied, then it is guaranteed that quantum advantage is achieved after a certain number of probes/uses.

Furthermore we have also designed an explicit protocol using TMSV states and a receiver based on photon counting and the maximum-likelihood estimation that allows us to beat the classical benchmark, in some cases after a smaller number of probes than the general quantum bound. Finally, we applied our study to some examples that are connected with thermal imaging and eavesdropper and additive-noise localization in different communication lines or among a sequence of frequencies. In conditions of low loss, we showed quantum advantage in various cases.

\smallskip
\textbf{Acknowledgments.}~This work was funded by the European Union via ``Quantum readout techniques and technologies'' (QUARTET, Grant agreement No 862644) and via Continuous Variable Quantum Communications” (CiViQ, Grant agreement No 820466), and by the EPSRC via the Quantum Communications Hub (Grants No. EP/M013472/1 and No. EP/T001011/1). Q.Z. acknowledges support from the Office of Naval Research under Grant Number N00014-19-1-2189 and the University of Arizona.

\appendix
\section{Behaviour of the classical fidelity function}\label{appendix:fidelity}
We now prove the statement in Section~\ref{sec:quantum advantage} that $F^{\mathrm{class}}_{\mathrm{loss/amp}}\to F^{\infty}_{\mathrm{loss/amp}}$ as $\tau\to 0$. Substituting $\tau=0$ into Eq.~(\ref{eq:fid_class_thermal}), we get
\begin{align}
&F_{\mathrm{loss/amp}}^{\mathrm{class}}(0,\epsilon_T,\epsilon_B)=\frac{\sqrt{\gamma_0+\delta_0}+\sqrt{\gamma_0-\delta_0}}{\delta_0},\\
&\gamma_0=4\epsilon_T\epsilon_B+1,\\
&\delta_0=2\left(\epsilon_T+\epsilon_T\right).
\end{align}
Rearranging, we get
\begin{align}
F_{\mathrm{loss/amp}}^{\mathrm{class}}(0,\epsilon_T,\epsilon_B)=\frac{\sqrt{2\gamma_0+2\sqrt{\gamma_0^2-\delta_0^2}}}{\delta_0},
\end{align}
and then, using
\begin{align}
\sqrt{\gamma_0\pm\delta_0}=\sqrt{(2\epsilon_T\pm1)(2\epsilon_B\pm1)},
\end{align}
we get
\begin{align}
\sqrt{\gamma_0^2-\delta_0^2}=\sqrt{(4\epsilon_T^2-1)(4\epsilon_B^2-1)}.
\end{align}
Thus, we have
\begin{align}
F_{\mathrm{loss/amp}}^{\mathrm{class},\tau=0}&=\frac{\sqrt{4\epsilon_T\epsilon_B+1+\sqrt{(4\epsilon_T^2-1)(4\epsilon_B^2-1)}}}{\sqrt{2}(\epsilon_T+\epsilon_B)}\\
&=F_{\mathrm{loss/amp}}^{\infty}.
\end{align}

The proofs that $\frac{dF}{d\tau}$ is positive semidefinite in the range $0\leq\tau<1$, that $\frac{dF}{d\tau}$ is negative semidefinite in the range $\tau>1$, and that $\frac{(F^{\mathrm{class}})^2}{F^{\infty}}\leq1$ for $\tau=\frac{1}{2}$ are given in the supplementary Mathematica files.

\end{document}